\documentclass[twoside,draft,reqno,12pt]{amsart}
\usepackage[english]{babel}
\usepackage{amsmath}
\usepackage{amssymb}
\usepackage{enumerate}

\usepackage{amsfonts}
\usepackage{graphicx}

\usepackage[ citecolor=black, linkcolor=black,
colorlinks, hypertexnames=false, plainpages=false]{hyperref}

\usepackage{xcolor}

\newtheorem{lemma}     {Lemma}[section]
\newtheorem{theorem}   [lemma]{Theorem}
\newtheorem{teorema1}   [lemma]{Theorem}
\newtheorem{prop}       [lemma]{Proposition}
\newtheorem{corollary}       [lemma]{Corollary}
\newtheorem{cong1}      [lemma]{Conjecture}
\newtheorem{remark}    [lemma]{Remark}
\newtheorem{definition}      [lemma]{Definition}

\numberwithin{equation}{section}

\newcommand{\und}{\underline}

\newcommand{\1}{\text{\bf 1}}    
\newcommand{\R}{\mathbb R}

\newcommand{\dis}{\displaystyle}

\newcommand{\mmmintone}[1]{{\dis{\int\kern -.38cm
-}}_{\kern-.21cm\substack{#1}}\;\;}
\newcommand{\mmmintwo}[2]{{\dis{\int\kern -.43cm
-}}_{\kern-.21cm\substack{#1}}^{\substack{#2}}\;\;}
\newcommand{\submint}{{\scriptstyle{\int\kern -.66em -}}}
\newcommand{\submintone}[1]{{\scriptstyle{\int\kern -.66em
-}}_{\scriptscriptstyle{\kern-.21em\substack{#1}}}}
\newcommand{\fracmint}{{\textstyle{\int\kern -.88em -}}}
\newcommand{\fracmintone}[1]{{\textstyle{\int\kern -.88em
-}}_{\scriptscriptstyle{\kern-.21em\substack{#1}}}\;}

\newcommand{\eps}{\epsilon}

\newcommand{\ga}{\gamma}
\newcommand{\Ga}{\Gamma}

\newcommand{\La}{\Lambda}

\newcommand{\nada}[1]{}

\def\be{\begin{equation}}
\def\ee{\end{equation}}

\def\supp{{\rm supp}\,}

\def\La{\Lambda}

\def\GG{\mathcal{G}}
\def\CC{\mathcal{C}}
\def\BB{\mathcal{B}}
\def\TT{\mathcal{T}}

\newcommand{\VV}{\mathcal{V}}
\def\II{\mathcal{I}}

\begin{document}
\today

\vskip.5cm
\title[Finite volume corrections and decay of correlations]
{Finite volume corrections and decay of correlations in the canonical ensemble}

\author{Elena Pulvirenti}
\address{Elena Pulvirenti,
Mathematical Institute, Leiden University, 
\indent P.O. Box 9512, 2300 RA Leiden, The Netherlands}
\email{pulvirentie@math.leidenuniv.nl}

\author{Dimitrios Tsagkarogiannis}
\address{Dimitrios Tsagkarogiannis,
Department of Mathematics, University of Sussex,
\indent Brighton BN1 9QH, UK}
\email{D.Tsagkarogiannis@sussex.ac.uk}

\begin{abstract}
We consider a classical system of $N$ particles confined in a box $\Lambda\subset\R^d$ 
interacting via a finite range pair potential. 
Given the validity of the cluster expansion in the canonical ensemble we 
compute the error between the finite and the infinite volume free energy and estimate it
to be bounded by the area of the surface of the box's boundary over its volume.
We also compute the truncated two-point correlation function
and find that the contribution from the ideal gas case is of the order $1/N$ 
while the contribution of the interactions is of the order $1/|\La|$ plus an exponentially small error with the distance.
\end{abstract}

\maketitle


\section{Introduction}
\label{intro}

A common practise of mathematical methods in physics is to consider
idealized situations by taking limits of the number of particles and/or the volume
of the system to infinity to study the limiting
thermodynamic quantities. 
However, in many situations of both theoretical (e.g. in the construction of coarse-grained Hamiltonians as in
the Lebowitz - Penrose theory) and practical use
(when dealing with realistic systems and computer simulations)
one is also interested in obtaining exact estimates of the error between
the infinite and the finite volume version of these quantities.
It is a longstanding problem to compute the error terms between the logarithm
of the (canonical or grand canonical) partition function and the limiting pressure or free energy. 
In the special case of the validity of the cluster expansion such questions
can be answered; see for example \cite{K}, \cite{C}, \cite{U} and the references therein, for some cases valid mainly
for lattice systems or the grand canonical ensemble.
In the present paper we work in the context of 
continuous systems in the canonical ensemble
and we calculate the
finite volume corrections to the free energy for both cases of periodic and zero (or general)
boundary conditions and estimate the relevant error.
The main technical tool is the cluster expansion whose validity has been established in
a previous work \cite{PT}. However, as it will be explained, its implementation for answering
the above question still requires to overcome several technical issues.
Moreover, as another application of the validity of the cluster expansion we also
investigate the decay of the two-point correlation function as the distance between
two particles increases.

The structure of the paper is as follows: in Section \ref{sec:2} we present the model and the results. Then, for completeness of the presentation, in Section \ref{sec:3} we give the basic ideas of
the cluster expansion in the canonical ensemble.
In the same section we also give the outline of the proof
of the finite volume corrections explaining why it is not a direct application of the existing
cluster expansion result and instead a new, more involved expansion has to be devised.
We present the proof in the subsequent three sections. In Section \ref{sec:5} we develop the new
version of the cluster expansion considering as polymers
{\it rooted} subsets of the set of labels of the particles carrying the extra information
whether the root is close to the boundary or not.
We apply the general theorem of cluster expansion for the new polymers and
conclude the proof in two steps. We first show in Section \ref{sec:6} that the finite volume error for the
polymers not vanishing in the thermodynamic limit is of the order of the boundary divided by the volume.
Second, in Section \ref{starstar}, we prove that the remaining (thermodynamically irrelevant) terms 
give a lower order contribution.
Last, in Section \ref{sec:7}, we give the proof of the decay of the two-point correlation function.

\bigskip


\section{The model and the results}
\label{sec:2}
We consider a configuration $ \mathbf{q}:=\{q_1,\ldots,q_N\}$ 
of $N$ particles (where $q_i$ is the position of the $i^{th}$ particle) confined in a box 
$\La(\ell):=(-\frac{\ell}{2},\frac{\ell}{2}]^d
\subset\mathbb{R}^d$ (for some $\ell>0$), which we also denote by $\La$ when we do not
need to explicit the dependence on $\ell$.
The particles interact via a pair potential
$V:\R^d\to\R\cup\{\infty\}$, which is stable and of finite range. Stability means that
 there exists $B\geq0$ such that:
\be\label{3}
\sum_{1\leq i<j \leq N} V(q_i-q_j) \geq -BN,
\ee
for all $N$ and all $q_1,...,q_N$. Finite range ($R>0$) means that
$V(q_i-q_j)= 0$ if $ |q_i-q_j| >R$,
where $|q_i-q_j|$ denotes the euclidean distance between two particles at positions $q_i$ and $q_j$. 
The requirement of finite range is only for technical reasons and will be clear in the sequel.
However, a similar result should be true under the hypothesis of temperedness:
\be\label{temper}
C(\beta):= \int_{\mathbb R^d} |e^{-\beta V(q)}-1| dq <\infty,
\ee
but the proof is more
involved and it is beyond the scope of the present paper.
In the finite-range potential case here, \eqref{temper} always holds because of \eqref{3} and we denote it by $C(\beta, R)$ in order to explicit the dependence on $R$.
A typical example of pair potential with the above features is the hard-core interaction
given by:
\be\label{hc}
V^{\text{hc}}(q_i-q_j)= 
\begin{cases}
+\infty, \qquad \text{if } |q_i-q_j| \leq R\\
0, \qquad \text{if } |q_i-q_j| > R
\end{cases}
\ee
Note that in this case $C(\beta,R)$ would be the volume of 
the $d$-dimensional ball with center $0$ and radius $R$, denoted by $B_{R}(0)$ (independent of $\beta$).

\bigskip

In the case of periodic boundary conditions, the {\it canonical partition function} of the system 
is given by
\be\label{1a}
Z^{per}_{\beta,\La,N}
:=\frac{1}{N!}\int_{\La^N} dq_1\,\ldots dq_N \,e^{-\beta H^{per}_{\La}(\mathbf q)},
\ee
where $H^{per}_{\La}$ is the energy of the system: 
\be\label{2}
H^{per}_{\La}( {\mathbf q})=
\sum_{1\leq i<j \leq N} V^{per}(q_i,q_j).
\ee
The potential $V^{per}$ captures the periodic boundary conditions and for
the case of finite range interaction it is given by
\be\label{0}
V^{per}(q_i,q_j):=
 \sum_{\substack{n=(n_1,\ldots,n_d) \\ n_i \in\{-1,0,1\}}} V(q_i-q_j+n\ell).
\ee

We also consider the case of \emph{zero boundary conditions}, i.e., outside the domain $\La$ there are no particles,
which is described by the Hamiltonian: $H_{\La}( {\mathbf q})=
\sum_{1\leq i<j \leq N} V(q_i,q_j)$. We denote by $Z^0_{\beta,\La,N}$ the corresponding partition function.
Given $\rho>0$ we define the {\it thermodynamic free energy} by
\be\label{4.1}
f_{\beta}(\rho):=\lim_{|\La|,N\to\infty,\,N=\lfloor\rho|\La|\rfloor} f_{\beta,\La}(N)
,\,\,\,{\rm where}
\,\,\,
f_{\beta,\La}(N)
:=
-\frac{1}{\beta|\La|}\log Z_{\beta,\La,N},
\ee
where  $|\La|=\ell^d$ is the volume of $\La$.
It is a general result (see \cite{FL}) that the above limit is independent of the boundary conditions, hence \eqref{4.1} holds for
both $Z^{per}_{\beta,\La,N}$ and $Z^0_{\beta,\La,N}$. 
We use the notation $Z_{\beta,\La,N}$ whenever the choice of boundary conditions is not relevant and
we avoid to specify it.
In \cite{PT}, for the case of periodic boundary conditions, using the cluster expansion method
it has been proved that
for the case of small enough densities, namely $\rho C(\beta)<c_0$ for some $c_0:= c_0(\beta,B)>0$,
the limit exists and it takes the explicit form:
\be\label{f}
\beta f_{\beta}(\rho)=\rho(\log\rho-1)-\sum_{n\geq 1}  \frac{1}{n+1}\beta_n\rho^{n+1}.
\ee
Here
$\beta_n$
is given by
\be\label{n4}
\beta_{n}:=
\frac{1}{n!}
\sum_{\substack{g\in\BB_{n+1}\\ V(g)\ni\{1\}}}\int_{\R^{dn}}\prod_{\{i,j\}\in E(g)}(e^{-\beta V(q_i-q_j)}-1)
dq_2\ldots dq_{n+1},\quad q_1:= 0,
\ee
where $\BB_{n+1}$ is the set of $2$-connected 
graphs $g$ on $(n+1)$ vertices
and $E(g)$ is the set of edges of the graph $g$.
We define a $2$-connected graph to be 
a connected graph which by removing each single vertex and all incident edges remains
connected. 

In the present paper we want to prove a more delicate estimate, namely
to calculate the terms which contribute to the finite volume corrections of the free energy
and estimate them by
the area of the surface $\partial\La$ of $\La$ divided by the volume of $\La$.
We denote by $|\cdot|$ the Lebesgue measure of either the surface or of the volume. 
\begin{theorem}\label{theoB1}
There exists a constant $c'_0:= c'_0(\beta,B)>0$, independent of $N$ and $\La$, 
such that if $\rho\,C(\beta,R)<c'_0$, there exist  constants $\tilde C(\rho), \hat C(\rho)>0$ for which:
\be\label{finitevol0}
\Big | \frac{1}{|\La|} \log Z^{per}_{\beta,\La,N} 
- \beta f_{\beta}(\rho) \Big| \leq
\tilde C(\rho) \frac{1}{|\La|},
\ee
for the case of {\it periodic} boundary conditions and
\be\label{finitevol}
\Big | \frac{1}{|\La|} \log Z^{0}_{\beta,\La,N} 
- \beta f_{\beta}(\rho) \Big| \leq
\hat C(\rho) \frac{|\partial \La|}{|\La|},
\ee
for the case of {\it zero} (general) boundary conditions.
Note also that $\rho=\frac{N}{|\La|}$
and $f_{\beta}(\rho)$ is given in \eqref{f}.
\end{theorem}

The proof of Theorem \ref{theoB1} will be outlined in Subsection \ref{proof}
and detailed in Sections \ref{sec:5}, \ref{sec:6} and \ref{starstar}.

\bigskip

Another byproduct of the cluster expansion is an expression for the truncated correlation functions.

\begin{definition}\label{def1}
Given a point $q_1\in\Lambda$, the one-point correlation function is given by
\begin{equation}\label{onepoint}
\rho^{(1)}_{\Lambda,N}(q_1):=\frac{1}{(N-1)!}\int_{\La^{N-1}} dq_2\ldots dq_N \frac{1}{Z_{\beta,\Lambda,N}}e^{-\beta H_{\Lambda}(q)}.
\end{equation}
Similarly, for $q_1,q_2 \in \La$, the two-point correlation function is given by
\begin{equation}\label{twopoint}
\rho^{(2)}_{\Lambda,N}(q_1,q_2):=\frac{1}{(N-2)!}\int_{\La^{N-2}} dq_3\ldots dq_N \frac{1}{Z_{\beta,\Lambda,N}}e^{-\beta H_{\Lambda}(q)}.
\end{equation}

\end{definition}

\medskip

Note that $\frac 1N\rho^{(1)}_{\Lambda,N}(q_1) dq_1$ is the probability of having {\it any} particle in 
a volume $dq_1$ at position $q_1$ (among $N$ particles). 
Indeed, if we consider periodic boundary conditions we have that
\begin{equation*}
\int_{\Lambda} \rho^{(1)}_{\Lambda,N}(q_1) dq_1
=N.
\end{equation*}
Similarly, defining $g^{(2)}_{\Lambda,N}(q):=\frac{1}{\rho^2}\rho^{(2)}_{\Lambda,N}(0,q)$ we can interpret the quantity
$\frac{1}{N-1}\rho g^{(2)}_{\Lambda,N}(q) dq$ 
as the probability of observing a second particle in a volume $dq$ at position $q$ given that there is
already a particle at the origin $0$. Note that using periodic boundary conditions we have that
\begin{equation*}
\int_{\Lambda} \rho g^{(2)}_{\Lambda,N}(q) dq 
=N(N-1)\frac{1}{\rho}\frac{1}{N!}\int_{\Lambda}dq
\int_{\La^{N-2}} dq_3\ldots dq_N \frac{1}{Z^{per}_{\beta,\Lambda,N}}e^{-\beta H^{per}_{\Lambda}(q)}
=N-1.
\end{equation*}

\medskip

\begin{remark}\label{rmk1}
In the case of the canonical ensemble, the two-point correlation function
does not factorize into the product of the one-point correlation functions, not even in the case of non-interacting particles. 
In fact, for the ideal gas we have that
\begin{equation}\label{corunlabelled}
\rho_{\Lambda,N}^{(2)}(q_1,q_2)-\rho_{\Lambda,N}^{(1)}(q_1)\rho_{\Lambda,N}^{(1)}(q_2)=
\frac{N(N-1)}{|\Lambda|^2}-\left(\frac{N}{|\Lambda|}\right)^2=\frac{N}{|\Lambda|^2},
\end{equation}
which indicates that we can not do better than $1/N$.
This is due to the fact that, having fixed a particle of the $N$ many, the choice of a second one is among the
remaning $N-1$.
However, this is not any more the case if we label the particles 
(defining a correlation function without the indistinguishability factor $N!$)
and
ask what is the probability that we find particle 2 at some position if
we know that particle 1 is already somewhere. 
But, on top of that, we still
have to take into account the interaction between the particles.
This is a separate issue which will be studied using
the ``labelled" correlation functions, as defined in \eqref{corrlab}.
Then we put together both issues and calculate the overall decay of correlations estimate in Corollary \ref{thmunlab}.
\end{remark}

\medskip

We view the canonical partition function
$Z_{\beta,\La,N}$ as a perturbation around the ideal case (where there are no interactions). Hence,
normalizing the measure by multiplying and dividing by $|\La|^N$ in \eqref{1a},
we write
\be\label{5.1}
Z_{\beta,\La,N}= Z_{\La,N}^{ideal} Z_{\beta,\La,N}^{int},
\ee
where 
\be\label{5.2}
Z_{\La,N}^{ideal}:=\frac{|\La|^N}{N!}
\quad \text{and}\quad
Z_{\beta,\La,N}^{int}:=
\int_{\La^N} \frac{dq_1}{|\La|}\,\ldots \frac{dq_N}{|\La|} \,e^{-\beta H_{\La}(\bf q)}.
\ee

Following Remark \ref{rmk1}, 
we define the \emph{labelled} $k$-point correlation functions as follows:

\begin{definition}
For each $k=1,2,\ldots,N$,
define
\begin{equation}\label{corrlab}
\rho^{(k),lab}_{\Lambda,N}(q_1,\ldots,q_k):=\int_{\La^{N-k}} \frac{dq_{k+1}}
{|\Lambda|}\ldots \frac{dq_N}{|\Lambda|} \frac{1}{Z^{int}_{\beta,\Lambda,N}}e^{-\beta H_{\Lambda}(q)},
\end{equation}
again without specifying the boundary conditions.
\end{definition}

\medskip

For the case of the labelled correlation functions, we obtain the following theorem:
\medskip

\begin{theorem}\label{thmlab}
Let $q_1$ and $q_2$ be two fixed positions in the domain $\Lambda$. Then there exist four 
constants $c_0, C_1, C_2, C_3>0$, independent on $\La$ and $N$, such that, for $\rho C(\beta,R)<c_0$,
we have:
\begin{equation}\label{cor1}
|\rho^{(1), lab}_{\Lambda,N}(q_1)|\leq C_1
\end{equation}
and
\begin{equation}\label{corr}
|\rho^{(2),lab}_{\Lambda,N}(q_1,q_2)-\rho^{(1),lab}_{\Lambda,N}(q_1)\rho^{(1),lab}_{\Lambda,N}(q_2)|
\leq 
\mathbf 1_{|q_1-q_2|\leq R}
\frac{1}{1- \frac{|C(\beta,R)|}{|\Lambda|}}
+C_2 \frac{C(\beta,R)}{|\La|}+
C_3 e^{-R^{-1}|q_1-q_2|} .
\end{equation}
Here $\rho^{(2),lab}_{\Lambda,N}$ and $\rho^{(1),lab}_{\Lambda,N}$ are given in \eqref{corrlab}.
\end{theorem}

\medskip
The proof will be given in Section \ref{sec:7}.
Note that for $N=2$ and for the case of hard cores of radius $R$ we can calculate the two-point truncated correlation functions
directly from the definitions and obtain:
\begin{eqnarray}\label{N2}
\rho^{(2),lab}_{\Lambda,2}(q_1,q_2)-\rho^{(1),lab}_{\Lambda,2}(q_1)\rho^{(1),lab}_{\Lambda,2}(q_2) & = &
\frac{|\Lambda|}{|\Lambda\setminus B_R(0)|}\mathbf 1_{|q_1-q_2| > R}-1\nonumber\\
&=&-\mathbf 1_{|q_1-q_2|\leq R}\frac{|\Lambda|}{|\Lambda\setminus B_R(0)|}
+\frac{|B_R(0)|}{|\Lambda\setminus B_R(0)|}
\end{eqnarray}
which are the first two terms of \eqref{corr}.
As a corollary we also obtain the case of the unlabelled correlation functions:

\medskip
\begin{corollary}\label{thmunlab}
Let $q_1$ and $q_2$ be two fixed positions in the domain $\Lambda$. Then 
there exist positive constants
$C_2,C_3$ and $C'$, such that
\begin{eqnarray*}
&& |\rho^{(2)}_{\Lambda,N}(q_1,q_2)-\rho^{(1)}_{\Lambda,N}(q_1)\rho^{(1)}_{\Lambda,N}(q_2)| \leq
\nonumber\\
&&
\left(\frac{N}{|\Lambda|}\right)^2 \Big[\mathbf 1_{|q_1-q_2|\leq R}
\Big(\frac{1}{1- \frac{|C(\beta,R)|}{|\Lambda|}}\Big)
+C_2 \frac{C(\beta,R)}{|\La|}
+C_3 e^{-R^{-1}|q_1-q_2|} \Big]
+C'\frac 1N \left( \frac{N}{|\Lambda|}\right)^2, 
\end{eqnarray*}
where $\rho^{(2)}_{\Lambda,N}$ and $\rho^{(1)}_{\Lambda,N}$ are given in \eqref{twopoint}
and \eqref{onepoint}.
\end{corollary}
\medskip

{\it Proof.}
We have:
\begin{align}\label{nothing}
\rho^{(2)}_{\Lambda,N}&
(q_1,q_2)-\rho^{(1)}_{\Lambda,N}(q_1)\rho^{(1)}_{\Lambda,N}(q_2) \nonumber\\
 &  
=\frac{N(N-1)}{|\Lambda|^2}
\rho^{(2),lab}_{\Lambda,N}(q_1,q_2)-\left(\frac{N}{|\Lambda|}\right)^2
\rho^{(1),lab}_{\Lambda,N}(q_1)\rho^{(1),lab}_{\Lambda,N}(q_2)\nonumber\\
& 
=\left(\frac{N}{|\Lambda|}\right)^2
\left(
\rho^{(2),lab}_{\Lambda,N}(q_1,q_2)-\rho^{(1),lab}_{\Lambda,N}(q_1)\rho^{(1),lab}_{\Lambda,N}(q_2)
\right)-
\frac{N}{|\Lambda|^2}
\rho^{(2),lab}_{\Lambda,N}(q_1,q_2).
\end{align}
The first term is bounded as in \eqref{corr}, while for the second 
combining \eqref{cor1} and \eqref{corr} we have:
\begin{equation}\label{constant}
|\rho^{(2),lab}_{\Lambda,N}(q_1,q_2)|\leq 
|\rho^{(2),lab}_{\Lambda,N}(q_1,q_2)-\rho^{(1),lab}_{\Lambda,N}(q_1)\rho^{(1),lab}_{\Lambda,N}(q_2)|
+|\rho^{(1),lab}_{\Lambda,N}(q_1)\rho^{(1),lab}_{\Lambda,N}(q_2)|
\leq C_2+C_3+C_1^2.
\end{equation}
This concludes the proof by choosing  $C'=C_2+C_3+C_1^2$.\qed

\bigskip


\section{Cluster expansion and strategy of proof of Theorem \ref{theoB1}}
\label{sec:3}

In this section 
we briefly recall the cluster expansion in the canonical ensemble
as proved in \cite{PT}. Elements of it will be used in the proofs of both
Theorem \ref{theoB1} and Theorem \ref{thmlab}.
In the second part of this section, we give the strategy for the proof of Theorem \ref{theoB1}.
The full proof will be given in Sections \ref{sec:5}, \ref{sec:6} and \ref{starstar}.

\subsection{Cluster expansion in the canonical ensemble}\label{clusterexp}

Note that what follows holds for both periodic and zero boundary conditions, so for simplicity we do not distinguish 
between the two cases. We will distinguish them once this becomes relevant.
For $Z_{\beta,\La,N}^{int}$ (see \eqref{5.2}) we use the idea of Mayer in \cite{MM} 
which consists of developing $e^{-\beta H_{\La}(\bf q)}$ in the following
way
\be\label{bo}
e^{-\beta H_{\La}(\bf q)}=\prod_{1\leq i<j\leq N}(1+f_{i,j})
=\sum_{E\subset\mathcal E(N)}\prod_{\{i,j\}\in E}
f_{i,j},
\ee
where $\mathcal E(N):=\{\{i,j\}:\, i,j\in[N],\,i\neq j\}$, $[N]:=\{1,...,N\}$
and
\be\label{fij}
f_{i,j}:=e^{-\beta V(q_i-q_j)}-1, 
\ee
for the case of zero  boundary conditions. For periodic boundary conditions, 
the cluster expansion is the same, replacing $f_{i,j}$ by
$f_{i,j}:=e^{-\beta V^{per}(q_i-q_j)}-1$.
Note that in the last sum in equation \eqref{bo} we have also the term with $E=\varnothing$ which gives $1$.

A {\it graph} is a pair $g:= (V(g),E(g))$, where $V(g)$
is the set of {\it vertices} and  $E(g)$ is the set of  {\it edges}, with $E(g) \subset \{U\subset V(g) : |U|=2 \}$, $|\cdot|$ denoting the {\it cardinality} of a set.
A graph $g=(V(g),E(g))$ is said to be {\it connected}, if for every pair $A,B\subset V(g)$
such that $A\cup B= V(g)$ and 
$A\cap B=\varnothing$, there is an edge $e\in E(g)$ such that $e\cap A\neq\varnothing$
and $e\cap B\neq\varnothing$.
Singletons are considered to be connected.
We use $\CC_V$ to denote the set of connected graphs on the set of vertices 
$V\subset [N]$, where we use the notation $[N]:=\{1,...,N\}$.

Two sets $V,V' \subset [N]$ are called {\it compatible} (denoted by $V\sim V'$) if $V\cap V'=\varnothing$;
otherwise we call them {\it incompatible} ($\nsim$). This definition induces in a natural way the notion of compatibility between graphs with set of vertices $V(g), V(g') \subset [N]$, i.e., $g\sim g'$ 
if  $V(g)\cap V(g')=\varnothing$.

With these definitions, 
to every set $E$ in equation \eqref{bo} we can associate a graph, i.e., a pair $g:= (V(g),E(g))$, where $V(g):=\{i:\,\exists e\in E \mbox{ with } i\in e\}\subset [N]$ and $E(g)=E$.
Note that the resulting graph does not contain isolated vertices.
It can be viewed as
the pairwise compatible (non-ordered) collection of its connected components,
i.e., $g:=\{g_1,\ldots,g_k\}_{\sim}$ for some $k$, 
where each $g_l$, $l=1,\ldots,k$, belongs to the set of all connected graphs on at most $N$ vertices and contains at least two vertices.
Hence,
\be\label{bo1}
e^{-\beta H_{\La}(\bf q)}=
\sum_{\substack{\{g_1,...,g_k\}_{\sim}\\ g_l \text{ connected}}} \prod_{l=1}^{k}
\prod_{\{i,j\}\in E(g_l)} f_{i,j},
\ee
where again the empty collection $\{g_1,...,g_k\}_{\sim}=\varnothing$ contributes the term $1$ in the sum.
Therefore, observing that integrals over compatible components factorize, we get 

\be\label{6}
Z_{\beta,\La,N}^{int}:=\sum_{\substack{\{g_1,...,g_k\}_{\sim}\\ g_l \text{ connected}}}
\prod_{l=1}^{k}\tilde\omega_{\La}(g_l)
=\sum_{\substack{\{V_1,...,V_k\}_{\sim}\\ |V_l|\geq 2,\,\forall l
}} \prod_{l=1}^{k}\omega_{\La}(V_l),
\ee
where
\be\label{7}
\tilde\omega_{\La}(g):=
\int_{\La^{|V(g)|}} \prod_{i\in V(g)} \frac{dq_i}{|\La|}  \prod_{\{i,j\}\in E(g)} f_{i,j}\quad \text{and} \quad
\omega_{\La}(V):= \sum_{g \in \CC_{V}}\tilde\omega_{\La}(g).
\ee

\medskip
An {\it abstract polymer model} ($\Ga$, $\mathbb{G}_{\Ga}$, $\omega$) 
consists of (i) a set of polymers 
$\Ga:=\{\ga_1,..., \ga_{|\Ga|}\}$, (ii) a binary symmetric relation $\sim$ of compatibility 
between the polymers (i.e., on $\Ga \times \Ga$), which is recorded into the
compatibility graph $\mathbb{G}_{\Ga}$
(the graph with vertex set $\Ga$ and with an edge between two polymers $\ga_i,\ga_j$ if and only if they are incompatible),
and (iii) a weight function $\omega : \Ga \to \mathbb{C}$.
Then, 
we have the following formal relation, which will become
rigorous by Theorem~\ref{thm2} below (see \cite{KP}, \cite{BZ} and \cite{NOZ}):
\be\label{poly1}
Z_{\Ga,\omega}:=\sum_{\{\ga_1,...,\ga_n\}_{\sim}} \prod_{i=1}^{n} \omega (\ga_i)=\exp\left\{
\sum_{ I\in \mathcal I} c_{ I}\omega^{ I}
\right\},
\ee
where
\be\label{7.2}
c_I=\frac{1}{I!}\sum_{G\subset\GG_I}(-1)^{|E(G)|},
\ee
or equivalently (\cite{BZ},\cite{D})
\be\label{7.21}
c_I= \frac{1}{I!} \frac{\partial^{\sum_{\ga}I(\ga)} \log Z_{\Ga,\omega}}{\partial^{I(\ga_1)} \omega(\ga_1) \cdots \partial^{I(\ga_n)} \omega(\ga_n)} \Big|_{\omega(\ga)=0}.
\ee

The sum in \eqref{poly1} 
is over the set $\mathcal I$ of all multi-indices $I:\Ga \to\{0,1,\ldots\}$,
 $\omega^I:=\prod_{\ga}\omega(\ga)^{I(\ga)}$,
 and, denoting
 $\supp I:=\{\ga \in \Ga :\,I(\ga)>0\}$,
 $\GG_I$ is the graph with $\sum_{\ga\in\supp I} I(\ga)$ vertices induced from
 $\GG_{\supp I} \subset \mathbb{G}_{\Ga}$ by replacing each vertex $\ga$ by the complete graph on
 $I(\ga)$ vertices.
  Furthermore, the sum in \eqref{7.2} is over all connected subgraphs $G$ of $\GG_I$ spanning the whole set of vertices of $\GG_I$
 and $I!:=\prod_{\ga\in\supp I} I(\ga)!$.
Note that if $I$ is such that  $\GG_{\supp I}$ is not connected (i.e., $I$ is not a \emph{cluster}) then
 $c_{I}=0$.

We state the general theorem as a slightly simplified version of \cite{BZ} and \cite{NOZ}.

\medskip

\begin{theorem}[Cluster Expansion]\label{thm2}
Assume that there are two 
non-negative functions $a,c:\Ga\to\R$ such that for every $\ga\in\Ga$,
$|\omega(\ga)|e^{a(\ga)+c(\ga)}\leq\delta$ 
holds, for some $\delta\in (0,1)$.
Moreover, assume that for each polymer $\ga'$
\be\label{7.4}
\sum_{\ga\nsim\ga'}|\omega(\ga)|e^{a(\ga)+c(\ga)}\leq a(\ga').
\ee
Then, for every polymer $\ga'\in\Ga$, we obtain that
\be\label{7.5}
\sum_{I:\,I(\ga')\geq 1} |c_I\omega^I| e^{\sum_{\ga\in\supp I}I(\ga) c(\ga)}\leq |\omega(\ga')| e^{a(\ga')+c(\ga')},
\ee 
where the coefficients $c_I$ are given in \eqref{7.21}.
\end{theorem}
  
\medskip

  \bigskip
In view of \eqref{6} we represent the partition function $Z_{\beta,\La,N}^{int}$
as a 
polymer model on 
$\mathcal V(N):=\{V:\,V\subset\{1,\ldots,N\},\,|V|\geq 2\}$
with weights $\omega_{\La}$ as in \eqref{7} and compatibility
graph $\mathbb G_{\mathcal V(N)}$.
Hence the abstract polymer formulation is given by the space
 ($\mathcal V(N)$,  $\mathbb{G}_{\mathcal V(N)}$, $\omega_{\La}$) and we are able 
 to check the convergence condition \eqref{7.4} 
 (see \cite{PT}) and thus
 write the following expansion:
 \be\label{expansion}
 Z^{int}_{\beta,\La,N} :=\sum_{\{ V_1,..., V_n\}_{\sim}} \prod_{i=1}^{n} \omega_{\La} ( V_i)=\exp\left\{
\sum_{ I\in \mathcal I} c_{ I}\omega_{\La}^{ I}
\right\},
 \ee
 where $\mathcal I$ is the set of all multi-indices $I: \mathcal V(N) \to\{0,1,\ldots\}$
 and the series is absolutely convergent.

\bigskip


\subsection{Strategy of the proof of Theorem \ref{theoB1}}\label{proof}

In this subsection we prove \eqref{finitevol0} and explain the strategy of the proof of \eqref{finitevol}.
We first recall the result of \cite{PT}. For the case of periodic boundary conditions, it has been proved that:

\begin{theorem}\label{thm0}
There exists a constant $c_0:= c_0(\beta,B)>0$, independent of $N$ and $\La$, 
and functions $F_{\beta,N,\La}(n)$, $n\in\mathbb N$,
such that if $\rho\,C(\beta)<c_0$ 
then 
\be\label{p3.1}
\frac{1}{|\La|}\log Z^{per}_{\beta,\La,N}=
\frac{1}{|\La|}\log\frac{|\La|^N}{N!}+
\frac{N}{|\La|}
\sum_{n\geq 1} F_{\beta,N,\La}(n),
\ee
with $N=\lfloor\rho|\La|\rfloor$.
In the thermodynamic limit
\be\label{n3}
\lim_{N,|\La|\to\infty,\,N=\lfloor\rho|\La|\rfloor}F_{\beta,N,\La}(n)=
\frac{1}{n+1}\beta_n\rho^{n},
\ee
for all $n\geq 1$,
where $\beta_n$ is defined in \eqref{n4}.
Furthermore, 
the functions $F_{\beta,N,\La}(n)$, $n\geq 1$, are given by
\be\label{p8}
F_{\beta,N,\La}(n)=\frac{1}{n+1}
\binom{N-1}{n}
\sum_{I:\,A(I)=[n+1]}
c_{I}\zeta_{\La}^{I}=\frac{1}{n+1}P_{N,|\La|}(n)B_{\beta,\La}(n),
\ee
with $A(I):=\cup_{V\in \supp I} V\subset \{1,...,N\}$,
\be\label{p9}
P_{N,|\La|}(n):=
\frac{(N-1)\ldots(N-n)}{|\La|^n}\quad\text{and}\quad
B_{\beta,\La}(n):=\frac{|\La|^n}{n!}\sum_{I:\, A(I)=[n+1]}
c_{I}\zeta_{\La}^{I}.
\ee
Moreover, there exist constants $C, c>0$ such that, for every $N$ and $\La$,
\be\label{n1}
|F_{\beta,N,\La}(n)|\leq C e^{-cn}.
\ee
\end{theorem}

\medskip

From the above theorem we then calculate the limit in \eqref{4.1} by applying the Dominated
Convergence Theorem combining the limit \eqref{n3} and the bound \eqref{n1}.
Given \eqref{expansion},\eqref{p8} and \eqref{p9}, we can write:
\begin{equation}\label{p0}
\frac{1}{|\La|} \log Z^{int}_{\beta,\La,N} =
\frac{1}{|\La|}\sum_{I\in\mathcal I} c_I \omega_{\La}^I
=\frac{N}{|\La|}
\sum_{n\geq 1}\frac{1}{n+1} P_{N,|\La|}(n)B_{\beta,\La}(n).
\end{equation}
We define by
\be\label{Bstar}
B^*_{\beta,\La}(n):=\frac{|\La|^n}{n!}\sum_{\substack{I\in \II^*:\\A(I)=[n+1]}}
c_{I}\omega_{\La}^{I}
\ee
the part of the sum $B_{\beta,\La}(n)$ restricted to multi-indices satisfying the following conditions:
\begin{eqnarray}
&& I(V)=1,\,\forall V\in\supp I,\,\,\, \text{and}\label{p1.3}\\
&& n+1= \sum_{V\in\supp I}(|V|-1)+1\label{p1.4}
\end{eqnarray}
and we denote by $\II^*$ the corresponding set.
In \cite{PT} it has been proved that under periodic boundary conditions
\be\label{periodic}
B^*_{\beta,\La}(n)=\frac{1}{|\La|}\frac{1}{n!}
\sum_{g\in \mathcal B_{n+1}}
\int_{\La^{n+1}}dq_1\ldots dq_{n+1}
  \prod_{\{i,j\}\in E(g)}
f_{i,j}
\ee
and consequently that
\be\label{p1.5}
\lim_{\La\to\mathbb{R}^d} B_{\beta,\La}(n)=\lim_{\La\to\mathbb{R}^d} B^*_{\beta,\La}(n)=\beta_n.
\ee
Furthermore, note that in our case of interactions with compact support and periodic boundary conditions 
it is easy
to see that actually 
$B^*_{\beta,\La}(n)=\beta_n$ for all $\Lambda$.
Then to prove \eqref{finitevol0} and \eqref{finitevol} one would split as follows:
\begin{align}\label{start}
\Big | \frac{1}{|\La|} \log Z_{\beta,\La,N} 
- \beta f_{\beta}(\rho) \Big| &\leq
 \Big| \frac{1}{|\La|} \log \frac{|\La|^N}{N!} - \rho(\log \rho - 1)\Big|  \\
&+\Big | \frac{1}{|\La|}\sum_{I\in\II^*} c_{I}\omega_{\La}^{I}-\sum_{n\geq 1} \frac{1}{n+1} \beta_n \rho^{n+1}  \Big|
+ \Big | \frac{1}{|\La|}\sum_{I\in\II^{**}} c_{I}\omega_{\La}^{I} \Big |
\end{align}
where by $\II^{**}:=\II\setminus \II^*$ we denote the remaining terms.
For the first term we use Stirling's approximation:
\be\label{stirling}
 \Big| \frac{1}{|\La|} \log \frac{|\La|^N}{N!} - \rho(\log \rho - 1)\Big| 
= \frac{1}{|\La|} \Big( \log \sqrt{2\pi N} + \frac1{12 N} + 0(N^{-3}) \Big).
\ee
For the third contribution a counting of the powers of $|\La|$ appearing in the numerator and
denominator (see \cite{PT}) shows
that each term is of order 
$\frac{1}{|\Lambda|}$ (or higher). However, we still have an infinite sum
and this will be addressed in Section \ref{starstar}, see \eqref{mainestimateI4}.
On the other hand, for the second term we have to estimate the error between the
finite volume integrals appearing in $\omega_{\La}^I$ and their infinite volume version in $f_{\beta}$.
Under periodic boundary conditions relation \eqref{periodic} holds and such a comparison is straightforward
(as it was mentioned after \eqref{p1.5}).
Hence, we only need to estimate the difference between $P_{N,|\La|}$ given in \eqref{p9} and $\rho^n$:
\begin{equation}\label{est1}
 \sum_{n\geq 1} \frac{1}{n+1} \rho^{n+1} \Big|\frac{N(N-1)\ldots(N-n)}{N^{n+1}}  -1 \Big| | \beta_n
|.
\end{equation}
We have:
\be
N(N-1)\ldots(N-n) = N^{n+1}\Big[
\Big(1- \frac{1}{N} \Big) \Big( 1- \frac{2}{N} \Big) 
\cdots \Big(1- \frac{n}{N}\big) \Big].
\ee
We first consider the case $n\leq  N^{1/2}$.
Since $\log(1-\frac iN)<0$ and for $x<0$ it is implied that $x\leq e^x-1<0$ we obtain
\be\label{one}
\Big| \prod_{i=1}^n(1-\frac{i}{N})-1 \Big|
\leq 
\sum_{i=1}^n |\log(1-\frac iN)|
\leq
c \sum_{i=1}^n \frac iN = c \frac{n(n+1)}{2N}.
\ee
In the last inequality we use the fact that for $x\leq\frac{1}{\sqrt 2}$ (since $n\leq  N^{1/2}$)
there exists $c \geq \sqrt 2$ such that $0>\log(1-x)>-cx$ (the latter is true for $c>\frac{1}{1-x}$).
Then for the case $n\leq  N^{1/2}$ we obtain the bound
\be\label{daje}
\frac{c}{N} \sum_{n\leq N^{1/2}} \frac{1}{n+1} \rho^{n+1} \frac{n(n+1)}{2}
|\beta_n|
\leq \frac{c'}{N},
\ee
for some $c'>0$.
On the other hand, for the case $n>  N^{1/2}$, we bound \eqref{est1} by
\be\label{two}
2\sum_{n\geq N^{1/2}} \frac{1}{n+1}
\rho^{n+1}
|\beta_n|
\leq 2 \rho \sum_{n\geq N^{1/2}} e^{-cn}
\leq 2 \rho e^{-c N^{1/2}},
\ee
since \eqref{n1} and \eqref{n3}
imply that $|\frac{1}{n+1}\rho^{n}\beta_n|\leq C e^{-cn}$.
This, together with \eqref{mainestimateI4}, proves
\eqref{finitevol0}.

On the other hand, for zero (or general) boundary conditions one would need
to split each integral in $\omega_{\La}^I$ into an interior and a boundary part. Then the
collection of all interior parts should be compared with the infinite volume free energy $f_{\beta}(\rho)$
(as it happens in the case of periodic boundary conditions).
But, to collect all these interior parts we need to 
rewrite the sum in \eqref{Bstar} as a sum over graphs (using \eqref{7}) which unfortunately is not convergent.
The remedy comes from a new cluster expansion where 
the information about whether we have an interior or a boundary integral 
is included in the definition of polymers. Thus, the new polymers consist of the sets of labels like before, 
plus some additional information
on whether all involved particles are far from the boundary of the box or not.
In this way, the desired bound for the contribution of polymers ``touching" the boundary of the box
comes for free as a corollary of the cluster expansion
theorem (see Proposition \ref{l2}).

\bigskip


\section{Cluster expansion on the space of rooted sets}
\label{sec:5}

To implement the new cluster expansion we follow Section \ref{sec:3} until  \eqref{6} and then
use the following splitting:
\be\label{partition}
1= \sum_{i \in V} \frac{1}{|V|} \Big( \1_{d(q_i,\La^c) <R|V|} +
\1_{d(q_i,\La^c) \geq R|V|} \Big)
\ee
for every $V\subset \{1,\ldots,N\}$ and 
where $d(\cdot,\cdot)$ is the Euclidean distance.
Inserting it in \eqref{6} we obtain:
\begin{align}
Z_{\beta,\La,N}^{int}=
\sum_{\substack{\{V_1,...,V_n\}_{\sim}\\ |V_l|\geq 2,\,\forall l
}} \prod_{l=1}^{n} \sum_{g\in\CC_{V_l}} &
\int_{\La^{|V_l|}} \frac{dq_1}{|\La|} \cdots \frac{dq_{|V_l|}}{|\La|}  \prod_{\{j,k\}\in E(g)} f_{j,k}
\sum_{i \in V_l} \frac{1}{|V_l|} \times \nonumber\\
&
\Big( \1_{d(q_i,\La^c) <R|V_l|} +
\1_{d(q_i,\La^c) \geq R|V_l|} \Big),
\end{align}
where again $f_{i,j}$ has two different possible definitions in the case of zero or periodic boundary conditions 
(see the discussion after \eqref{fij}).
Given $V_l$ and $i\in V_l$ the quantity in the parenthesis represents the two cases: 
either the particle with label $i$ is closer to the boundary than $R|V_l|$ giving a boundary contribution or not. 
We introduce a parameter $\eps_i\in\{0,1\}$ to distinguish between these two cases.
We consider a real function $F$ defined as follows:
\be\label{Feps}
F(\eps_i):= (1-\eps_i)\, \1_{d(q_i,\La^c) <R|V_l|} + \eps_i \, \1_{d(q_i,\La^c) \geq R|V_l|}.
\ee
Then,
\begin{align}
&
F(0)= \1_{d(q_i,\La^c) <R|V_l|}\\
&
F(1)= \1_{d(q_i,\La^c) \geq R|V_l|}
\end{align}
and hence:
\begin{equation}\label{part}
Z_{\beta,\La,N}^{int}=
\sum_{\substack{\{V_1,...,V_n\}_{\sim}\\ |V_l|\geq 2,\,\forall l
}} \prod_{l=1}^{n} \sum_{i \in V_l} \sum_{\eps_i=0,1}
\sum_{g\in\CC_{V_l}} 
\int_{\La^{|V_l|}}\prod_{k\in V_l} \frac{dq_k}{|\La|}   \prod_{\{j,k\}\in E(g)} f_{j,k}
 \frac{1}{|V_l|} F(\eps_i).
\end{equation}

We unify the sums over $V$, $i$ and $\epsilon_i$
by defining the polymers of the new expansion to be the triplets $(V,i,\eps_i)$, where $V\in\VV(N)$, 
$\VV(N):=\{V: V \subset [N] \}$,
$i\in V$ and $\eps_i\in\{0,1\}$. One may think of many copies of a set of vertices $V$, as many as the
number of its elements (choice of $i$) 
each one taken two times (as $\eps_i$ can take the values $0$ or $1$).
The new polymers differ from the old ones as they are \emph{rooted} sets, with the label $i$ being 
the root and {\it coloured}, where $\epsilon_i\in\{0,1\}$ are the two colours.
We use the notation $\und{V}$ to indicate the triplet $(V,i,\eps_i)$ and we refer to $V$ as the
support of $\und{V}$: $V=\supp \und{V}$.
Two polymers $\und{V}_1$ and $\und{V}_2$ are  compatible 
 if  their supports are compatible (see Section \ref{sec:3}).

With slight abuse of notation we still define the activity as a function $\omega_{\La}: \VV(N) \times \{1,...,N\} \times \{0,1\} \to \mathbb{R}$
and for a polymer $\und V=(V,i,\eps_i)$ it has the following expression
\be
\omega_{\La}(\und V)= \sum_{g \in \CC_{V}} \int_{\La^{|V|}} \frac{dq_1}{|\La|} \cdots \frac{dq_{|V|}}{|\La|}
 \prod_{\{j,k\}\in E(g)} f_{j,k}
 \frac{1}{|V|} F(\eps_i).
\ee
Then the canonical partition function \eqref{part} can be written as:
\be\label{ceU1}
Z_{\beta,\La,N}^{int}=
\sum_{\substack{\{\und{V}_1,...\und{V}_k\}_{\sim} \\ |\supp \und{V}_i| \geq 2}} \prod_{l=1}^{k} 
\omega_{\La}(\und V_l).
\ee
Thus, we are again in the context of Theorem \ref{thm2} and obtain
\be\label{Ss}
Z_{\beta,\La,N}^{int}=
\exp\Big\{
\mathcal S_{\beta,\La,N}^{(0)}+\mathcal S_{\beta,\La,N}^{(1)}
\Big\},
\qquad
\text{where, for $i=0,1$},
\,\,\,
\mathcal S_{\beta,\La,N}^{(i)}:=
\sum^{(i)}_{I} c_{I}\omega_{\La}^{I}.
\ee
The sum $\sum^{(0)}$ contains all multi-indices $I$ such that
there is at least one choice of $V\in \VV(N)$ and $i\in V$ with $\eps_i=0$
(boundary contributions).
On the other hand,
the sum $\sum^{(1)}$ contains only those multi-indices for which
for \emph{every} choice of $\und V\in \supp I$ the value of the corresponding $\eps_i$ is equal to $1$,
i.e., it consists of the polymers which are localized in the interior of $\Lambda$.

With this new cluster expansion, following the arguments
of subsection \ref{proof},
we refine \eqref{start}
by first splitting between the terms $\mathcal S_{\beta,\La,N}^{(0)}$ which interact with the boundary 
and those ($\mathcal S_{\beta,\La,N}^{(1)}$) which are far from it.
Then, the latter we split as in \eqref{start} between the ones that produce the irreducible
coefficients ($*$ terms) and the rest ($**$ terms). Hence we have:
\be\label{next}
\log Z_{\beta,\La,N}^{0, \, int}=
\mathcal S_{\beta,\La,N}^{(0)}+\mathcal S_{\beta,\La,N}^{(1),*}+ \mathcal S_{\beta,\La,N}^{(1),**},
\ee
where $\mathcal S_{\beta,\La,N}^{(0)}$, $\mathcal S_{\beta,\La,N}^{(1),*}$
and $\mathcal S_{\beta,\La,N}^{(1),**}$ are given in \eqref{Ss} taking into account 
the further restrictions $*$ and
$**$.
Summing up we have:
\begin{align}
&\Big| \frac{1}{|\La|} \log Z^0_{\beta,\La,N} - \beta f_{\beta}(\rho) \Big| =\notag\\
&\Big| \frac{1}{|\La|}  \log \frac{|\La|^N}{N!} +  \frac{1}{|\La|} \Big( \mathcal S_{\beta,\La,N}^{(0)}
+ \mathcal S_{\beta,\La,N}^{(1),*} + \mathcal S_{\beta,\La,N}^{(1),**} \Big) 
- \rho(\log \rho-1) - \sum_{n\geq 1} \frac{1}{n+1} \beta_n \rho^{n+1} \Big|.
\end{align}
Using \eqref{stirling}, we need to estimate the following quantities:
\begin{align}\label{i1}
& \mathcal I_1:= \Big| \frac{1}{|\La|} \mathcal S_{\beta,\La,N}^{(0)} \Big| \\ \label{i3}
& \mathcal I_2:= \Big|  \frac{1}{|\La|} \mathcal S_{\beta,\La,N}^{(1),*} -  \sum_{n\geq 1} \frac{1}{n+1} \beta_n \rho^{n+1} \Big| \\ \label{i4}
& \mathcal I_3:= \Big| \frac{1}{|\La|} \mathcal S_{\beta,\La,N}^{(1),**}\Big|. 
\end{align}

For the first term (which is the main contribution) see Proposition \ref{l2} below.
The terms $\mathcal I_2$ and $\mathcal I_3$ will be treated in Sections \ref{sec:6} 
and \ref{starstar},
respectively.

In what follows we first check the validity of the hypothesis of Theorem \ref{thm2}
and then based on \eqref{7.5} we derive a bound on $\mathcal I_1$.

\vspace{0.5cm}
We have, exactly as in \cite{PT}:

\begin{lemma}\label{KP}
There exists a constant $c'_0:= c'_0(\beta,B)>0$ such that for $\rho \, C(\beta,R)< c'_0$
there exist positive constants $a$ and $\delta \in (0,1)$
such that for every $\La:=(-\frac{\ell}{2},\frac{\ell}{2}]^d
\subset\mathbb{R}^d$:
\be\label{conv}
\sup_{\und V:= (V,i,\eps_i)}
|\omega_{\La}(\und V)|e^{a|\und V|}\leq\delta
\ee
holds, where $N=\lfloor\rho |\La|\rfloor$.
Moreover, for every $\und{\tilde V} $:
\be\label{c0}
\sum_{ \und V \nsim \und{\tilde V}} |\omega_{\La}(\und V)| 
e^{a|\und V|} \leq a|\und{\tilde V}|.
\ee

\end{lemma}

\vspace{0.5cm}

{\it Proof}.
To bound $|\omega_{\La}(\und V)|$ 
we use the tree-graph inequality (see the original references \cite{P}, \cite{B}; here we use 
the particular form given in \cite{PU}, Proposition 6.1 (a)):
\be\label{tgi}
\Big|\sum_{g \in \CC_{n}} \prod_{\{j,k\} \in E(g)} f_{j,k}\Big| \leq 
e^{2\beta B n}
\sum_{T \in\TT_n} \prod_{\{j,k\}\in E(T)} |f_{j,k}|,
\ee
where $\TT_n$ and $\CC_{n}$  are respectively the set of trees and connected graphs with $n$ vertices.
For a fixed $\und V=(V,i,\eps_i)$ with $|\und V|=|V|=n$, we have
\begin{equation}\label{c1}
|\omega_{\La}(\und V)| e^{a |\und V|} 
\leq 
e^{(a+2\beta B) n} \sum_{T \in\TT_n} \int_{\La^n} \frac{dq_1}{|\La|} \cdots \frac{dq_n}{|\La|} \prod_{\{j,k\}\in E(T)} |f_{j,k}|
\,\frac{1}{n} |F(\eps_i)|.
\end{equation}

Given a rooted tree $T$ let us call $(a_1,b_1),(a_2,b_2),...,(a_{n-1},b_{n-1})$ its edges.
We consider $1$ as the root of the tree and  using the change of variables:
\be
y_k= q_{a_k}-q_{b_k}, \quad \forall k=2,...,n, \qquad
y_1=q_1.
\ee
We obtain:
\begin{align} \label{big}
\int_{\La^n} \frac{dq_{1}}{|\La|} \cdots \frac{dq_{n}}{|\La|} \prod_{\{j,k\} \in E(T)} | f_{j,k} |  &=
\frac{1}{{|\La|}^n}\int_{\La^n} dq_{1}\cdots dq_{n} \prod_{k=1}^{n-1} | f_{a_k,b_k} |  \,|F(\eps_i)|\notag \\
&
\leq \frac{1}{{|\La|}^n}\int_{\La} dq_{1} \int_{\La} dy_2\cdots  \int_{\La} dy_{n} \prod_{k=2}^{n} | e^{ -\beta V(y_k)}-1| 
\,|F(\eps_i)| \notag \\
&
\leq \frac{1}{{|\La|}^n}\int_{\La} dq_{1} \,|F(\eps_i)| \cdot \left[ \int_{\La} dx | e^{ -\beta V(x)}-1|\right]^{n-1}
\notag \\
& \leq C(\beta,R)^{n-1} \frac{1}{{|\La|}^n}\int_{\La} dq_{1} \,|F(\eps_i)| 
\leq  \frac{C(\beta,R)^{n-1}}{{|\La|}^{n-1}},
\end{align}
if we use the generic bound $|F(\eps_i)|\leq 1$ for both values of $\eps_i$ (see \eqref{Feps}).
Note that in the case $\epsilon_i=0$ we obtain the better bound:
\be
\leq \frac{C(\beta,R)^{n-1}}{{|\La|}^{n-1}} \,2 \,d\, R\, n\, \frac{1}{\ell},
\ee
because we integrate over an interior annulus of width $R n$ along $\partial \Lambda$.
Then, since the number of all trees in $\TT_n$ is $n^{n-2}$,
from \eqref{c1} we obtain  (recalling that $N=\lfloor\rho|\La|\rfloor$):
\be\label{acti}
|\omega_{\La}(\und V)| e^{a |\und V|} \leq e^{(2\beta B+a) n} 
\frac{ n^{n-2} }{|\La|^{n-1}} C(\beta,R)^{n-1}=
e^{(2\beta B+a) }\Big( e^{(2\beta B+a)} C(\beta,R) \rho \Big)^{n-1},
\ee
or the better estimate 
\be\label{acti2}
|\omega_{\La}(\und V)| e^{a |\und V|} \leq 
\frac{e^{(2\beta B+a) }}{n} \Big( e^{(2\beta B+a)} C(\beta,R) \rho \Big)^{n-1}  \,2 \, \frac{d\,R}{\ell},
\ee
in the case $\epsilon_i=0$.
If we choose $\rho\, C(\beta,R)$ such that: 
\be\label{c2.1}
\delta':=\rho e^{(2\beta B+a)}C(\beta,R)<1,
\ee
then for every $\und V$
\be
|\omega_{\La}(\und V)| e^{a |\und V|} \leq 
\frac{1}{2}\rho\, C(\beta,R) e^{2(2\beta B+a)},
\ee
by using the bound $2\leq n\leq N$ and the fact that 
$\rho e^{(2\beta B+a)}C(\beta,R)<1$. Then, defining 
$\delta:=
\rho\, C(\beta,R) \,e^{2(2\beta B+a)}  $,
\eqref{conv} is satisfied.

\vspace{0.5cm}

For every fixed $\und{\tilde V}$  we have:
\begin{equation}\label{c3}
\sum_{\und V \nsim \und{\tilde V}} |\omega_{\La}(\und V)| 
e^{a|\und V|} =
\sup_{k \in {\tilde V}} |\tilde V| \sum_{\und V:  V \ni k}
|\omega_{\La}(\und V)| 
e^{a|\und V|} 
=
\sup_{k \in {\tilde V}} |\tilde V|
\sum_{V: V \ni k} \sum_{i\in V}
 \sum_{\eps_i=0,1}
|\omega_{\La}(V, i,\epsilon_i)| e^{a |V|}   
\end{equation}
and if we sum over the cardinality of the set $V$, $V\ni k$:
\begin{align}
\sum_{n\geq 2} &\binom{N-1}{n-1} 2 n
e^{(2\beta B+a) n} 
\frac{ n^{n-2} }{|\La|^{n-1}} 
C(\beta,R)^{n-1} \nonumber \\ 
&\leq
2 e^{(2\beta B+a)} 
\sum_{n\geq 2}\Big(\frac{N}{|\La|}\Big)^{n-1} \frac{ n^{n-1} }{(n-1)!} (C(\beta,R) \, e^{(2\beta B+a)})^{n-1}
 \leq 
 2
 \frac{e^{(2\beta B+a)} 
  }{\sqrt{\pi}} 
\frac{\delta' e}{1-\delta' e}.
\end{align}
In the first expression we have used \eqref{acti} and estimated the sum over $i \in V$  by $n$, 
while in the last inequality we used Stirling's bound: $n!\geq n^n e^{-n}\sqrt{2\pi n}$ with $\rho C(\beta,R)$ small
such that $\delta' e <1$.
Choosing $a$ such that 
$ 2
\frac{e^{(2\beta B+a)} 
 }{\sqrt{\pi}} 
\frac{\delta' e}{1-\delta' e}<a$
(by taking $\delta'$ small enough)
we get the bound \eqref{c0}.
\qed

\vspace{0.5cm}

Then the estimate on $\mathcal I_1$ follows:

\begin{prop}\label{l2}
There exists $C>0$ such that
\begin{equation}\label{boundary}
\Big|
\frac{1}{|\Lambda|}\mathcal S_{\beta,\La,N}^{(0)}
\Big|
\leq C \frac{1}{\ell}, 
\end{equation}
\end{prop}

{\it Proof.}
From \eqref{7.5} applied to the clusters over the new polymers we have that
\begin{eqnarray}\label{KPcor1}
\Big|
\frac{1}{|\Lambda|}\mathcal S_{\beta,\La,N}^{(0)}
\Big| 
& \leq &
\frac{1}{|\Lambda|}\sum_{\substack{\und V_1: \\ \und V_1=(V,i,0)}}
\sum_{I:\, I(\und V_1)\geq 1} |c_I \omega_{\Lambda}^I |\nonumber\\
&\leq &
\frac{N}{|\Lambda|}
\sum_{V: V \ni 1} \sum_{i\in V}
|\omega_{\La}(V, i,0)| e^{a |V|}   \leq
C(\rho) \frac{1}{\ell}, 
\end{eqnarray}
using \eqref{acti2} and
where $C(\rho)$ is a positive constant which depends on $\rho=\frac{N}{|\Lambda|}$:
\be\label{Cofrho}
C(\rho):=
\sum_{n\geq 2} N^{n-1} \frac{1}{(n-1)!} 
e^{(2\beta B+a) n} 
\frac{ n^{n-2} }{|\La|^{n-1}} C(\beta,R)^{n-1}.
\ee
The sum is convergent and hence we
conclude the proof. \qed


\section{Estimate for $\mathcal I_2$}\label{sec:6}

Recalling the splitting \eqref{next} of the partition function with zero boundary conditions, we can repeat
exactly the same steps of the new cluster expansion  for the partition function
with periodic boundary conditions:
\be\label{next2}
\log Z_{\beta,\La,N}^{per, \, int}=
\mathcal S_{\beta,\La,N}^{(0), \,per}+\mathcal S_{\beta,\La,N}^{(1),*,\,per}+ \mathcal S_{\beta,\La,N}^{(1),**,\, per}.
\ee
Here $\mathcal S_{\beta,\La,N}^{(0),\, per}$, $\mathcal S_{\beta,\La,N}^{(1),*, \, per}$
and $\mathcal S_{\beta,\La,N}^{(1),**, \, per}$ are the corresponding 
to the splitting in \eqref{next} terms but computed with
periodic boundary conditions.
The key observation here is that
\begin{equation}\label{observe}
\mathcal S_{\beta,\La,N}^{(1),*}=\mathcal S_{\beta,\La,N}^{(1),*,\,per}.
\end{equation}
This is true since all the terms in the sum are in the interior of $\La$
(as it is indicated by the upper script $(1)$ in the sum).

Following the proof of Proposition \ref{l2} for the case of periodic boundary conditions we have that
there exists $C>0$ such that
\begin{equation}\label{boundaryper}
\Big|
\frac{1}{|\Lambda|}\mathcal S_{\beta,\La,N}^{(0),\, per}
\Big|
\leq \frac{C}{\ell}.
\end{equation}
Furthermore, repeating the steps leading to \eqref{mainestimateI42} we obtain that
\begin{equation}\label{I4per}
\Big| \frac{1}{|\La|} \mathcal S_{\beta,\La,N}^{(1),**,\, per}\Big|\leq \frac{C}{|\La|}. 
\end{equation}
Then, from \eqref{next2} using \eqref{observe}, \eqref{boundaryper} and \eqref{I4per} we get:
\begin{equation}\label{provisoire}
\Big|
\mathcal S_{\beta,\La,N}^{(1),*}-\log Z_{\beta,\La,N}^{per, \, int}
\Big|
\leq  \frac{C}{\ell}.
\end{equation}
Furthermore, comparing with \eqref{finitevol0} (using also \eqref{stirling}) we obtain that
\begin{equation}\label{I3}
|\mathcal I_2| \leq  \frac{C}{\ell}
\end{equation}
concluding the estimate for $\mathcal I_2$.

%
\bigskip

\section{Estimate for $\mathcal I_3$}\label{starstar}
In this section we prove that there exists a constant $C>0$ such that
\begin{equation}\label{mainestimateI4}
\Big|
\frac{1}{|\La|}\sum_{I\in \II^{**}} c_{I}\omega_{\La}^{I} \Big|
\leq C\frac{1}{|\La|},
\end{equation}
where $\II^{**}$ has been defined in \eqref{start}.
The bound \eqref{mainestimateI4} is a direct consequence of \eqref{bound} and of Lemmas \ref{lemI4_1} and \ref{lemI4_2} below.
The estimate for $\mathcal I_3$ is analogous
since the constraint $(1)$ in \eqref{i4} does not play any role and can be removed going to an upper bound.
Then it is enough to repeat the same steps as for the proof of \eqref{mainestimateI4}
since the main ingredient is again the validity of Theorem \ref{thm2}
(and in particular of \eqref{7.5})
which is true for both expansions over polymers $V$ and polymers $\und V$.
Thus, we also have that
\begin{equation}\label{mainestimateI42}
|\mathcal I_3|
\leq C\frac{1}{|\La|}.
\end{equation}
Note that the two estimates \eqref{mainestimateI4} and \eqref{mainestimateI42} are
valid for both periodic and zero boundary conditions.

We start by rewriting the sum over multi-indices as a sum over ordered sequences
$(V_1,\ldots, V_n)$. 
In order to do this, let us introduce the following truncated functions:
\be\label{U2}
\phi^T(V_1,...,V_n):=
\sum_{g\in\CC_n} \prod_{\{i,j\} \in E(g)} -\1_{\{V_i\nsim V_j\}},
\ee
if $n\geq 2$, while $\phi^T(V)=1$.
 We exploit the restriction to the set $\mathcal I^{**}$. This means that there exists a collection of, say
 $k$ elements, each one in $\mathcal V(N)$, for which the conditions \eqref{p1.3} and \eqref{p1.4} fail.
 Hence, given $k\leq n$ and the labels $i_1,\ldots, i_k \in\{1,\ldots,n\}$ 
 of these elements (all different), we define the following 
 sets in $\mathcal V(N)^n$
 :  for $k\geq 3$ we let
 \begin{eqnarray}\label{setA}
 A_{i_1,\ldots,i_k} & := & \left\{ \{V_1,\ldots,V_n\}\, \text{with}\,\,\{1,\ldots,n\}\supset\{i_1,\ldots,i_k\}:
\, 
 \exists \, v_1,\ldots, v_k\in [N], \,\text{all different} \right.\nonumber\\ 
&& \text{s.t.}\,
 V_{i_l}\cap V_{i_{l+1}} \supset \{v_l\}, \,\text{for}\,\, l=1,\ldots,k, \text{with} \, i_{k+1}:= i_1 \} 
 \end{eqnarray}
and for $k=2$
 \begin{equation}\label{Aij}
 A_{i_1,i_2}:=\{
 \{V_1,\ldots,V_n\}, \text{with}\,\,\{1,\ldots,n\}\supset\{i_1,i_2\}:
 |V_{i_1}\cap V_{i_2}|\geq 2\}.
 \end{equation}
 
 We have:
 \begin{eqnarray}\label{bound}
 \Big|
\frac{1}{|\La|}\sum_{I\in \II^{**}} c_{I}\omega_{\La}^{I} \Big|
 & \leq &
 \frac{1}{|\Lambda|}
\sum_{n\geq 2}\frac{1}{n!} \sum_{(V_1,\ldots,V_n)}\mathbf 1_{
\cup_{i_1,\ldots,i_k} A_{i_1,\ldots,i_k}}
|\phi^T({V}_1,...,{V}_n)|
 \prod_{l=1}^{n} |\omega_{\La}( V_l)|
 \nonumber\\
 & \leq &
  \frac{1}{|\Lambda|}
\sum_{n\geq 2}\frac{1}{n!}
 \sum_{(V_1,\ldots,V_n)}
 \sum_{(i_1,\ldots, i_k)} \mathbf 1_{A_{i_1,\ldots,i_k}} |\phi^T({V}_1,...,{V}_n)|
 \prod_{l=1}^{n} | \omega_{\La}( V_l)|
 \nonumber\\
 & \leq &
  \frac{1}{|\Lambda|}
 \frac{1}{k!}
 \sum_{(V_1,\ldots,V_k)}
 \mathbf 1_{A_{1,\ldots,k}}
 \prod_{l=1}^{k} | \omega_{\La}( V_l)|
 \cdot \nonumber\\
 &&
 \cdot\big(1+\sum_{n\geq k+1}\frac{1}{(n-k)!}
 \sum_{\substack{(V_{k+1},\ldots,V_n) \\ 
 }
 }
 | \phi^T({V}_1,...,{V}_n) |
 \prod_{l=k+1}^{n} |\omega_{\La}( V_l)|
 \big),
 \end{eqnarray}

where in the last inequality we use the fact that each choice of $i_1,\ldots,i_k$ is the same, hence we
consider one choice (on the first $k$ positions) and
multiply with the cardinality $\binom{n}{k}$.
We first prove that the sum over $n\geq k+1$ is bounded. 
Then the estimate that this is of order $\frac{1}{|\Lambda|}$ comes from the first sum over the
sets $V_1,\ldots,V_k$ with the constraint $A_{1,\ldots, k}$ (i.e., that are not satisfying \eqref{p1.3} and \eqref{p1.4}).
We have:

\vspace{0.5cm}

\begin{lemma}\label{lemI4_1}
Given $V_1,\ldots,V_k$ pairwise incompatible,
there exists a constant $C$ such that

 \be\label{duestelle2}
 \sum_{n\geq k+1}\frac{1}{(n-k)!}
 \sum_{(V_{k+1},\ldots,V_n)}
 | \phi^T({V}_1,...,{V}_n) |
 \prod_{l=k+1}^{n} | \omega_{\La}( V_l)|
 \leq C 
 \prod_{l=1}^k
 |V_l|  e^{|V_l|}.
 \ee

\end{lemma}

A similar result holds under the assumptions of
Theorem \ref{thm2} for some function $c$,
if we replace $\omega_{\La}( V_l)$ by
$\omega_{\La}( V_l)e^{c(V_l)}$.

\vspace{0.5cm}

 {\it Proof}.
To prove it we follow the argument given in Cammarota \cite{C}. 
 The main idea is to write $\phi^T(V_1,\ldots,V_n)$ as a sum over trees with branches emerging from each $V_i$, $i=1,\ldots,k$
 and then use the strategy leading to \eqref{7.5} by summing each branch independently.

We first work for $k=2$ and then argue that the general case $k\geq 3$ is similar.
Using the tree-graph inequality (see \cite{PU}, Proposition 6.1 (a))  we obtain:

 \begin{eqnarray}\label{W}
 \sum_{(V_3,\ldots,V_n)}
 | \phi^T({V}_1,...,{V}_n) |
 \prod_{l=3}^{n} | \omega_{\La}( V_l)|
&\leq &
\sum_{(V_3,\ldots,V_n)}
 \sum_{T\in\TT_n} \prod_{\{i,j\}\in E(T)} |-\mathbf 1_{\{V_i\nsim V_j\}}|
 \prod_{l=3}^{n} | \omega_{\La}( V_l)|\nonumber\\
&
=
&
\sum_{T \in \TT_n} \sum_{\substack{(V_{3},...,V_n)\\ \mathbb{G}( V_1,...,V_n)\supset T \cup \{1,2\}}}
\prod_{l=3}^{n} \big| \omega_{\La}( V_l) \big|.
\end{eqnarray}

We sum over $T\in \TT_n$ as follows:
we sum over two trees $T_1$ (respectively $T_2$) with root the label $1$ (respectively $2$) of cardinality $n_1+1$
(respectively $n_2+1$)
and an edge from every vertex of $T_1$ to the root $2$ of $T_2$. Note that $n_1+n_2=n-2$.
To implement it we first partition the set of labels $\{3,\ldots,N\}$ into two subsets $N_1$ and $N_2$ and construct two trees 
$T_1$ and $T_2$ from each one with the additional root $1$ and $2$. 
The two trees are connected by $\{\xi, 2\}$ for some $\xi\in T_1$.
We find an estimate by removing the extra edge $\{\xi,2\}$ as well as $\{1,2\}$ from the constraint in \eqref{W} and obtain
\be\label{crucial2}
\eqref{W}
\leq
\sum_{\substack{\{N_1,N_2\}\\\text{ part. of } \{3,...,n\}}} \sum_{\substack{T_1 \in \TT_{|N_1 \cup \{1\}|}\\ T_2 \in \TT_{|N_2 \cup \{2\}|}}}
\sum_{\xi\in V(T_1)}
\prod_{j=1,2}
 \sum_{\substack{(V_l)_{l\in N_j}\\ \mathbb{G}( V_l)_{l\in N_j\cup\{j\}}\supset T_j}}
\prod_{l\in N_j} \big| \omega_{\La}( V_l) \big|,
\ee
where we also recall that $V(T_1)$ is the set of vertices of tree $T_1$.

A similar relation is true for the general case $k\geq 3$. We decompose $T$ into a tree $T_1$
with root the label $1$, then an edge from some $\xi_1\in T_1$ to the root $2$ of the subtree $T_2$
and similarly another edge from some $\xi_2\in T_2$ to root $3$ of the next subtree $T_3$ etc.
We relax the constraint by removing the edges between the subtrees and obtain:
\be\label{crucial3}
\eqref{W}
\leq
\sum_{\substack{\{N_1,\ldots, N_k\}\\\text{ part. of } \{k+1,...,n\}}} \sum_{\substack{T_1 \in \TT_{|N_1 \cup \{1\}|}
\\ 
\cdots
\\
T_k\in\TT_{|N_k \cup \{k\}|}}}
\sum_{\xi_1\in V(T_1),\ldots, \xi_k\in V(T_k)}
\prod_{j=1}^k
 \sum_{\substack{(V_l)_{l\in N_j}\\ \mathbb{G}( V_l)_{l\in N_j\cup\{j\}}\supset T_j}}
\prod_{l\in N_j} \big| \omega_{\La}( V_l) \big|.
\ee

To calculate the right hand side of \eqref{crucial3}
we sum over $n_1,...,n_k$, $n_i$ being the cardinality of $N_i$, $i=1,...,k$:
\be\label{2tree}
\sum_{\substack{n_1,\ldots, n_k\\ n_1+\ldots+n_k=n-k}} \frac{(n-k)!}{n_1! \ldots n_k!}
(n_1+1)\ldots (n_k+1)
\prod_{j=1}^k
\Big( \sum_{T \in \TT_{n_j +1}} \mathcal {A}_{n_j,V_j}(T) \Big),
\ee
where for $T \in \TT_{n_j +1}$ with root $V_j$ we have defined the following quantity:
\be\label{mathcalA}
 \mathcal {A}_{n_j,V_j}(T) := \sum_{\substack{(V'_i)_{i=2,\ldots, n_j+1}\\ 
 \mathbb{G}( V_j,(V'_i)_{i=2,\ldots, n_j+1})\supset T}} 
\prod_{l=2}^{n_j+1}  \big| \omega_{\La}( V'_l) \big|.
\ee
Following \cite{C}, we have that for a tree $T\in\TT_m$, 
whose vertices $i \in \{1,...,m\}$ have degrees $d_i$,
the following estimate holds:
\be\label{2tree1}
\mathcal {A}_{m,V_1}(T) \leq |V_1|^{d_1} a^{m-1} \prod_{j\in\{2,...,m\}}(d_j-1)!,
\ee
where $a$ is the constant appearing in Lemma \ref{KP}.
Moreover, the number of trees on $m$ vertices with degrees $d_1,\ldots,d_m$ is given 
by the Cayley formula:
\be\label{cayley}
\frac{(m-2)!}{\prod_{i=1}^{m} (d_i-1)!}.
\ee
Thus, for the sums over trees in \eqref{2tree} using \eqref{2tree1} and \eqref{cayley} we have:
\begin{align}
\sum_{T \in \TT_{n_1 +1}} \mathcal {A}_{n_1,V_1}(T)& \leq
\sum_{\substack{d_1,...,d_{n_1+1}:\\ \sum_{i=1}^{n_1+1} d_i=2n_1}}
\frac{(n_1-1)!}{\prod_{i=1}^{n_1+1}(d_i-1)!} |V_1|^{d_1} a^{n_1} 
\prod_{i=2}^{n_1+1}(d_i-1)! \nonumber \\ 
&= a^{n_1} (n_1-1)! \sum_{d_1=1}^{n_1} \frac{ |V_1|^{d_1} }{(d_1-1)!}
\Ga_{n_1}(2n_1-d_1),
\end{align}
where
\be
\Ga_{k}(m):= \sum_{\substack{d_1,...,d_{k}: \\ d_1+...+d_{k}=m}}1.
\ee
$\Ga_{k}(m)$ are the Stirling numbers of the second kind, since they count the number of ways to partition $m$ labelled objects into $k$
 unlabelled subsets.
Applying an induction argument on $k$ we obtain that
\be
\Ga_{k}(m) \leq \frac{m^{k-1}}{(k-1)!},
\ee
thus,
\be
\Ga_{n_1}(2n_1-d_1)\leq
\frac{(2n_1-d_1)^{n_1-1}}{(n_1-1)!} \leq \frac{1}{2} (2e)^{n_1}.
\ee
Hence we have:
\be\label{2tree2}
\sum_{T \in \TT_{n_1 +1}} \mathcal {A}_{n_1,V_1}(T)\leq
\frac{1}{2} (2ae)^{n_1} (n_1-1)! |V_1|
\sum_{d_1=1}^{n_1} \frac{ |V_1|^{d_1-1} }{(d_1-1)!}
\ee
and therefore the l.h.s. of \eqref{duestelle2}, using \eqref{2tree} and \eqref{2tree2} is bounded by
\begin{align}
&\sum_{n\geq 3} \frac{1}{(n-k)!} 
\sum_{\substack{n_1,\ldots,n_2\\ n_1+\ldots+n_2=n-k}} \frac{(n-k)!}{n_1! \ldots n_k!}
\prod_{i=1}^k\Big[ \frac{1}{2} (2ae)^{n_i} (n_i-1)! |V_i|
e^{|V_i|} \Big] \nonumber \\ 
&\leq\prod_{i=1}^k 
\Big[
\sum_{n_i\geq 1} \frac{1}{n_i!}\frac{1}{2} (2ae)^{n_i} (n_i-1)! |V_i|
e^{|V_i|} 
\Big]\nonumber
\\ 
&\leq\prod_{i=1}^k 
\Big[
\sum_{n_i\geq 1} \frac{1}{2} (2ae)^{n_i} |V_i|
e^{|V_i|}
\Big] \nonumber\\ 
&\leq  C\prod_{i=1}^k
 |V_i|  e^{|V_i|},
\end{align}
which concludes the proof.
\qed

\bigskip
To obtain a bound for \eqref{bound} we exploit the fact that the fixed sets  $V_1,\ldots, V_k$
are in $A_{1,\ldots,k}$.
This is given in the following Lemma :

\begin{lemma}\label{lemI4_2}
For all sets $A_{1,\ldots,k}$ defined in \eqref{setA} we have that there exists a positive constant $C$ such that
\be\label{unosulvol}
\frac{1}{|\La|} 
\sum_{\{V_1,\ldots,V_k\}}
 \mathbf 1_{A_{1,\ldots,k}}
 \prod_{l=1}^{k} 
\left( | \omega_{\La}( V_l)|
 |V_l|  e^{|V_l|}
 \right)
 \leq C \frac{1}{|\La|}.
 \ee
\end{lemma}

\vspace{0.5cm}

 {\it Proof}.
 Let $n_1,\ldots,n_k$ be the cardinalities of $V_1,\ldots,V_k$.
 Choosing the labels for an element $V_1,\ldots,V_k$ of $A_{1,\ldots,k}$ 
 we look for the highest power of $N$. This occurs when
 we obtain a factor of $\binom{N}{n_1}\sim \frac{N^{n_1}}{n_1!}$ for the first, $\frac{N^{n_2-1}}{(n_2-1)!}$ for the second (since there is at least one common label
 with $V_1$), similarly for the next ones and for the last $\frac{N^{n_k-2}}{(n_k-2)!}$ since it has one common with
 the previous and another common with the first one (see the definition of $A_{1,\ldots,k}$ in \eqref{setA}).
 So the power of $N$ is at most $n_1+(n_2-1)+\ldots+(n_k-2)=\sum_{i=1}^k n_i -k$.
All other cases have a smaller exponent for $N$ since the sets $V_i$, $i=1,\ldots,k$
may share more labels (compare with \eqref{p1.3} and \eqref{p1.4}).
 On the other hand, from \eqref{acti}, the activities of each $V_i$ satisfy
\be\label{again}
 |\omega_{\La}(V_i)|  |V_i| e^{a |V_i|} \leq e^{(2\beta B+a) n_i} 
\frac{ {n_i}^{n_i-2} }{|\La|^{n_i-1}}  n_i C(\beta,R)^{n_i-1}.
\ee
Hence, the power of $|\Lambda|$ in the denominator is again $\sum_{i=1}^k n_i -k$ and combined with
the numerator gives $\rho$ to that power.
Thus, overall, all terms in the sum in \eqref{unosulvol} is of the order $\frac{1}{|\Lambda|}$ or higher.
Furthermore, the sum of all the terms is convergent because of \eqref{7.5}.
\qed

\vspace{0.5cm}


\section{Correlations}
\label{sec:7}

In this section we first show how the validity of the cluster expansion is related to 
the calculation of the truncated correlation function
and then conclude by
proving Theorem \ref{thmlab}.
Let $h_1,h_2$ be measurable functions on $\La$.
We define:
\begin{equation}\label{fcn}
\Psi_{\Lambda,N,h_1,h_2}(a_1,a_2):=\int_{\La^N} \frac{dq_1}{|\Lambda|}\ldots \frac{dq_N}{|\Lambda|}
\prod_{i=1,2}(1+a_i h_i(q_i))
e^{-\beta H_{\Lambda}(\bf q)}. 
\end{equation}
Given some fixed $q'_1, q'_2\in \La$,
we choose the functions $h_1,h_2$ such that 
\be\label{acca}
h_i(q_i):=\begin{cases}
1 \quad \text{if } q_i=q_i' \\
0 \quad \text{ otherwise},
\end{cases}
\ee
 for both $i=1,2$.
Then, with a slight abuse of notation, 
we denote by $\Psi_{\Lambda, N, q_1'}$ the corresponding function \eqref{fcn} for this special choice of $h_1$ 
and for $a_2= 0$ and we have:
\begin{equation}\label{trick1}
\rho^{(1),lab}_{\Lambda,N}(q_1')
= |\Lambda|\frac{\partial}{\partial a_1}\log \Psi_{\Lambda, N, q_1'}(0,0).
\end{equation}
Similarly, 
we call $\Psi_{\Lambda,N,q'_1,q'_2}(a_1,a_2)$
the corresponding function \eqref{fcn} with  $h_1$ and $h_2$ as before and
we have that:
\begin{equation}\label{trick}
\rho^{(2),lab}_{\Lambda,N}(q_1',q_2')-\rho^{(1),lab}_{\Lambda,N}(q_1')\rho^{(1),lab}_{\Lambda,N}(q_2')=
|\Lambda|^2
\frac{\partial^2}{\partial a_1\partial a_2}\log \Psi_{\Lambda, N,q'_1,q'_2}(0,0).
\end{equation}
Thus, to calculate \eqref{trick1} and \eqref{trick} we
first compute the cluster expansion for the function $\Psi_{\Lambda, N,h_1,h_2}(a_1,a_2)$,
then apply it for the two particular cases appearing in \eqref{trick} and \eqref{trick1}
and evaluate it at $a_1=0$ and $a_2=0$.

\medskip
We define the space $\mathcal V^*:=\{1\}\cup\{2\}\cup_{A\subset\{1,2\}}\mathcal V_A$,
where the elements of $\mathcal V_A$ are the ordered pairs $(V,A)$ for $V\in\mathcal V$ with $V\supset A$ and $A\subset\{1,2\}$.
We
define the activities
\begin{equation}\label{activities}
\omega_{\Lambda,h_1,h_2}((V,A)):=\prod_{i\in A}a_i\sum_{g\in\mathcal C_V}\int_{\La^{|V|}} 
\prod_{i\in A}h_i(q_i)  \prod_{\{i,j\}\in E(g)} f_{i,j}\prod_{i\in V} \frac{dq_i}{|\La|}.
\end{equation}
Note that for the special polymers $\{1\}$ and $\{2\}$ we have the following activities:
\begin{equation}\label{specialactivity}
\omega_{\Lambda,h_i}(\{i\}):=a_i \int_{\La} h_i(q_i)\frac{dq_i}{|\La|}, \qquad i=1,2.
\end{equation}
If $A=\varnothing$ we have that $\omega_{\Lambda,h_1,h_2}((V,\varnothing))=\omega_{\Lambda}(V)$ as
given in \eqref{7}.
Following the strategy of Section \ref{sec:3} we can write
\begin{equation}\label{abstract}
\Psi_{\Lambda,N,h_1,h_2}(a_1,a_2)=\sum_{\substack{\{(V_1,A_1),...,(V_k,A_k)\}_{\sim}\\ (V_l,A_l)\in\mathcal V^*
}}\prod_{l=1}^k \omega_{\Lambda,h_1,h_2}((V_l,A_l)).
\end{equation}
Moreover, we say that two elements are compatible, $(V_1,A_1)\sim(V_2, A_2)$, if $V_1\sim V_2$.
N.B.: In the sequel, for simplicity of the notation, we denote the elements of $\mathcal V^*$ either as pairs $(V,A)$ or
just $V$ if the explicit knowledge of $A$ is irrelevant, as for example in \eqref{cor}, \eqref{daniel}, \eqref{estimate2} and elsewhere.
We are in the context of the abstract polymer model
in the space $(\mathcal V^*, \mathbb G_{\mathcal V^*}, \omega_{\Lambda,h_1,h_2})$
and we apply Theorem \ref{thm2}. One can check the convergence condition
exactly as in Lemma \ref{KP}
(see also \cite{PT}), with the only modification
that now a given element $V$ 
has to be chosen among
the new augmented set
$\mathcal V^*$, which just gives an extra factor 4.
Then we obtain
\begin{equation}\label{clexp}
\log\Psi_{\Lambda,N,h_1,h_2}(a_1,a_2)=\sum_{I\in\mathcal I(\mathcal V^*)}c_I
\omega_{\Lambda,h_1,h_2}^I.
\end{equation}
Moreover, as in \eqref{7.5}, we have the bound:
\begin{equation}\label{cor}
\sum_{I:\,I(V')\geq 1} |c_I\omega_{\Lambda,h_1,h_2}^I| e^{\sum_{V\in\supp I}I(V) c(V)}\leq |\omega_{\Lambda,h_1,h_2}(V')| e^{a(V')+c(V')},
\end{equation}
for every $V'\in\mathcal V^*$ and for some non-negative functions $a,c:\mathcal V^*\to\mathbb R$.

\medskip

{\it Proof of Theorem \ref{thmlab}.}
For the proof of \eqref{corr}, 
we use \eqref{trick} and, in order to estimate its right hand side,
we have to identify the non vanishing terms. These are the ones of order $a_1 a_2$
(and of order $1/|\Lambda|^2$).
We denote them by $\mathcal I_{a_1,a_2}$
and divide them into the following two classes:

The first class, $\mathcal I_{a_1,a_2}^{(1)}$, is given by:
\begin{itemize}
\item The clusters with a polymer in $\mathcal V_{\{1\}}$ and the polymer $\{2\}$ 
(or the polymer $\{1\}$ and a polymer in $\mathcal V_{\{2\}}$), both with multiplicity one, and all others in 
$\mathcal V_{\varnothing}$ with any multiplicity. 

\item The clusters with a polymer in $\mathcal V_{\{1\}}$ and a polymer in $\mathcal V_{\{2\}}$, both with multiplicity one,
 and any other polymer in $\mathcal V_{\varnothing}$ with any multiplicity. 
\end{itemize}

In the first case we have the activities:
\begin{equation}\label{firstbound}
\omega_{\Lambda,h_1}((V,\{1\}))=a_1\sum_{g\in\mathcal C_V}\int_{\La^{|V|}} 
h_1(q_1)  \prod_{\{i,j\}\in E(g)} f_{i,j}\prod_{i\in V} \frac{dq_i}{|\La|},\qquad  \,\omega_{\Lambda,h_2}(\{2\})=a_2
\int_{\La} h_2(q_2) \frac{dq_2}{|\Lambda|}
\end{equation}
and similarly for $\omega_{\Lambda,h_2}((V,\{2\}))$ and $\omega_{\Lambda,h_1}(\{1\})$.
Using again the formulation with ordered sequences instead of multi-indices, as in Section \ref{starstar},
the contribution of the first case to the r.h.s. of \eqref{clexp} is:
\begin{eqnarray}\label{daniel}
 && 
 \sum_{V}
 |\omega_{\Lambda,h_1}((V,\{1\}))|
|\omega_{\Lambda,h_2}(\{2\})|
 \cdot\Big(1+\sum_{n\geq 3}\frac{1}{(n-2)!}
 \sum_{\substack{(V_{3},\ldots,V_n) }}
 | \phi^T({V}_1,...,{V}_n) |
 \prod_{l=3}^{n} |\omega_{\La}( V_l)|
 \Big) \nonumber\\
&& \leq 
 \sum_{V}
  |\omega_{\Lambda,h_1}((V,\{1\}))|
|\omega_{\Lambda,h_2}(\{2\})|
 (1+|V|e^{|V|+1})
\end{eqnarray}
where we used Lemma \ref{lemI4_1}. Note that the second sum is over polymers $V_3,...,V_n\in \mathcal V_{\varnothing}$ (there can be repetitions).
If $N=2$, we have $V=\{1,2\}$ and all polymers $V_3,\ldots, V_n$ are copies of $(\{1,2\},\emptyset)$ with
activity
$|\omega_{\Lambda}((\{1,2\},\emptyset))|\leq\frac{C(\beta,R)}{|\Lambda|}$. 
In this case for the special choice of $h_1$ we have that
 \be
 |\omega_{\Lambda,q'_1}((\{1,2\},\{1\}))|
\leq
\frac{a_1}{|\La|^2} C(\beta),
 \ee
which yields
\be\label{bound1}
 |\omega_{\Lambda,q'_1}((\{1,2\},\{1\}))|
| \omega_{\Lambda,q'_2}(\{2\})| (1+2e^{3})\leq
 a_1 a_2\frac{1}{|\La|^2}C \frac{C(\beta,R)}{|\La|}. 
 \ee 
 
For $N\geq 3$ we multiply and divide with $e^{R^{-1}|q_1'-q_2'|}$.
Noticing that
$|q_1'-q_2'|\leq R\sum_{V}(|V|-1)$,
we apply again Lemma \ref{lemI4_1}, but for the case of activities multiplied by $e^{c|V|}$, and obtain:

 \begin{eqnarray}\label{bound2}
 &&
 e^{-R^{-1}|q'_1-q'_2|}
 \sum_{V}
 \omega_{\Lambda,h_1}((V,\{1\})) e^{|V|}
\omega_{\Lambda,h_2}(\{2\})
\times\nonumber\\
&&
 \Big(1+\sum_{n\geq 3}\frac{1}{(n-2)!}
 \sum_{\substack{(V_{3},\ldots,V_n) }}
 | \phi^T({V}_1,...,{V}_n) |
 \prod_{l=3}^{n} |\omega_{\La}( V_l)|e^{c|V_l|}
 \Big)\nonumber\\
&&
\leq
e^{-R^{-1}|q'_1-q'_2|}  
\sum_{V}
|\omega_{\Lambda,q'_1}((V,\{1\}))|
|\omega_{\Lambda,q'_2}(\{2\})|
\Big(|V| e^{|V|+1}\Big) \nonumber\\
&&
\leq
\frac{1}{|\La|^2} a_1a_2e^{-R^{-1}|q'_1-q'_2|} 
\sum_{n\geq 2}\binom{N-1}{n-1} \frac{n^{n-2}}{|\La|^{n-1}} C(\beta,R)^{n-1} n e^{(2\beta B+4) n} 
\nonumber\\
&&
\leq\frac{1}{|\La|^2} a_1a_2 C e^{-R^{-1}|q'_1-q'_2|} 
\sum_{n\geq 2} \Big( \rho C(\beta,R) e^{2\beta B+4}\Big)^{n-1}\leq 
C' \frac{1}{|\La|^2} a_1a_2 e^{-R^{-1}|q'_1-q'_2|}, 
\end{eqnarray}
where $C'=C'(\rho,\beta,R)$.

The last case in $\mathcal I_{a_1,a_2}^{(1)}$ is when clusters contain polymers $(V_1,\{1\})$,
$(V_2,\{2\})$ for some $V_1$, $V_2$ and any $V_3,...,V_k\in\mathcal V$.
Similarly to \eqref{bound2} we have:
\begin{eqnarray}\label{estimate3}
&&
e^{-R^{-1}|q'_1-q'_2|} \sum_{\{V_1,V_2\}}|\omega_{\Lambda,q'_1}((V_1,\{1\}))|e^{|V_1|}
|\omega_{\Lambda,q'_2}((V_2,\{2\}))|e^{|V_2|}
(\prod_{l=1,2}|V_l|e^{|V_l|}) \leq
\nonumber\\
&&
\leq
a_1 a_2 \frac{1}{|\La|^2}e^{-R^{-1}|q'_1-q'_2|}
\sum_{n_1,n_2\geq 2}\binom{N-1}{n_1-1}\binom{N-n_1-2}{n_2-1}
 \frac{n_1^{n_1-1}}{|\La|^{n_1-1}} e^{(2\beta B +2)n_1}C(\beta,R)^{n_1-1}\times \nonumber\\
 &&
 \times \frac{n_2^{n_2-1}}{|\La|^{n_2-1}} e^{(2\beta B +2)n_2}C(\beta,R)^{n_2-1}
  \nonumber\\
&&
\leq
  a_1 a_2 \frac{1}{|\La|^2}e^{-R^{-1}|q'_1-q'_2|}
 \sum_{n_1\geq 2}  \Big( \rho C(\beta,R) e^{2\beta B+2}\Big)^{n_1-1}
  \sum_{n_2\geq 2}  \Big( \rho C(\beta,R) e^{2\beta B+2}\Big)^{n_2-1}
\end{eqnarray}

Putting together \eqref{bound1}, \eqref{bound2} and \eqref{estimate3} we get:
\be\label{proof1}
\Big|\sum_{I\in\mathcal I_{a_1,a_2}^{(1)}}c_I
\omega_{\La,q'_1,q'_2}^I
\Big|\leq C' \frac{1}{|\La|^2} a_1a_2 e^{-R^{-1}|q'_1-q'_2|} +\frac{1}{|\La|^2} a_1a_2
C \frac{C(\beta,R)}{|\La|}. 
\ee

The second class, denoted by $\mathcal I_{a_1,a_2}^{(2)}$, 
consists of clusters containing a polymer in $\mathcal V_{\{1,2\}}$
with multiplicity one
and all others in $\mathcal V_{\varnothing}$. 
For $N=2$, clusters contain the polymer $(\{1,2\},\{1,2\})$ with multiplicity one
and the polymer $(\{1,2\},\emptyset)$ with any multiplicity $n$ and we denote by
$c_{1,n}$ the corresponding multi-index. The two activities are:
\be
\omega_{\Lambda,q'_1,q'_2}((\{1,2\},\{1,2\}))=-a_1 a_2\frac{1}{|\La|^2}\mathbf 1_{|q'_1-q'_2|\leq R},
\qquad 
|\omega_{\Lambda}((\{1,2\},\emptyset))|\leq \frac{C(\beta,R)}{|\Lambda|}.
\ee
We have that
\be
c_{1,n}= \frac{1}{n!} \sum_{G\subset \mathcal G_{n+1}} (-1)^{|E(G)|}=
 \frac{n+1}{(n+1)!} \sum_{G\subset \mathcal G_{n+1}} (-1)^{|E(G)|}= (n+1) c_{n+1},
\ee
 where $c_{n+1}$ is the multi-index in the cluster expansion with only one polymer 
with multiplicity $n+1$. On the other hand, we know from expanding the logatithm
for a general activity, say $X$, that
\be
\log(1+X)=\sum_{n\geq 1}\frac{(-1)^{n+1}}{n} X^n=\sum_{n\geq 1}c_n X^n
\ee
and hence 
\be
c_n=\frac{(-1)^{n+1}}{n}.
\ee
This implies that $c_{1,n}=(-1)^{n+2}=(-1)^n$.
Thus,
\begin{eqnarray}\label{finally}
&&
\Big|\omega_{\Lambda,q'_1,q'_2}((\{1,2\},\{1,2\}))\Big|\Big(1+ \sum_{n\geq 1} 
\Big| c_{1,n} \omega_{\Lambda}((\{1,2\},\emptyset))\Big|^n\Big)\nonumber\\
&\leq&
\frac{a_1 a_2}{|\La|^2}\mathbf 1_{|q'_1-q'_2|\leq R} \Big(1+\frac{C(\beta,R)}{|\La|}+\Big(\frac{C(\beta,R)}{|\La|}\Big)^2+
\Big(\frac{C(\beta,R)}{|\La|}\Big)^3+...\Big)
\nonumber\\
&=&
\frac{a_1 a_2}{|\La|^2}\mathbf 1_{|q'_1-q'_2|\leq R}
\Big(\frac{1}{1- |C(\beta,R)|/|\Lambda|}\Big).
\end{eqnarray}

For the general case we have clusters consisting of the polymer 
$(V,\{1,2\})$ with multiplicity one and activity 
\begin{equation}\label{secondbound}
\omega_{\Lambda,h_1,h_2}((V,\{1,2\}))=a_1a_2\sum_{g\in\mathcal C_V}\int_{\La^{|V|}} 
\prod_{i=1,2} h_i(q_i)  \prod_{\{i,j\}\in E(g)} f_{i,j}\prod_{i\in V} \frac{dq_i}{|\La|},
\end{equation}
together with  any other polymer in $\mathcal V_{\varnothing}$ with any multiplicity.
Using the fact that $(|V|-1) R\geq |q_1'-q_2'|$,
we obtain:
\begin{equation}\label{estimate1}
\sum_{\substack{V\in\mathcal V:\\V\supset\{1,2\}} }
\sum_{\substack{I\in\mathcal I(\mathcal V^*): \\ I(V;\{1,2\})\geq 1}} |c_I \omega_{\Lambda,q_1',q_2'}^I| \leq
e^{-R^{-1}|q'_1-q'_2|}   \sum_{\substack{V\in\mathcal V:\\V\supset\{1,2\}} }
|\omega_{\Lambda,q'_1,q'_2}((V,\{1,2\}))| e^{|V|-1},
\end{equation}
which leads to the same bound as in \eqref{bound2}.
Hence for the class $\mathcal I_{a_1,a_2}^{(2)}$ we obtain
\be\label{proof2}
\Bigg|\sum_{I\in\mathcal I_{a_1,a_2}^{(2)}}c_I\omega_{\Lambda,q'_1,q'_2}^I\Bigg|
\leq
\frac{a_1 a_2}{|\La|^2}\mathbf 1_{|q'_1-q'_2|\leq R}
\Big(\frac{|\La|}{|\La|- |C(\beta,R)|}\Big)+
C' \frac{1}{|\La|^2} a_1a_2 e^{-R^{-1}|q'_1-q'_2|}.
\ee

We conclude the proof of  \eqref{corr} by putting together estimates 
\eqref{proof1} and \eqref{proof2} (without the factors $a_1$ and $a_2$).

\bigskip

To prove \eqref{cor1} we proceed in a similar fashion.
We want to compute the terms in \eqref{clexp} which will give non-zero contribution
to \eqref{trick1}. The terms which give contributions of first order in $a_1$ (and of order $|\La|$)
are those with only one element in $\mathcal V_{\{1\}}$ (with multiplicity one) and all others in $\mathcal V_{\varnothing}$.
We denote this set of multi-indices by $\mathcal I_{a_1}$.
For a generic polymer $(V,\{1\})\in\mathcal V_{\{1\}}$ we have activity
$\omega_{\Lambda,h_1}((V,\{1\}))$, as defined in \eqref{firstbound}.
Then, by applying \eqref{cor} we obtain that
\begin{equation}\label{estimate2}
\Bigg|\sum_{I\in\mathcal I_{a_1}}c_I\omega_{\Lambda,q'_1}^I\Bigg|
\leq
\sum_{V\in\mathcal V_{\{1\}}}
\sum_{I:\, I(V)\geq 1} |c_I \omega_{\Lambda,q'_1}^I|
\leq \sum_{V\ni 1} |\omega_{\Lambda,q'_1}(V)| e^{a(V)}\leq \frac{C}{|\La|},
\end{equation}
which concludes the proof for $C=C_1$.
\qed

\bigskip

{\bf Acknowledgments.}
It is a great pleasure to thank Errico Presutti for suggesting us the problem and for his continuous advising.
We also acknowledge discussions with Marzio Cassandro, Sabine Jansen and Thierry Bodineau who gave us intuition about the estimate of Theorem \ref{thmlab}.
Moreover, we are indebted to one of the referees for the careful revision of the manuscript
and the detailed comments for the improvement of the presentation.
The research of both authors was partially supported by the FP7-REGPOT-2009-1 project
``Archimedes Center for Modeling, Analysis and Computation" (under grant agreement no 245749).
E. P. is further supported by ERC Advanced Grant 267356 VARIS of Frank den Hollander.


\end{document}